# Mirror-enhanced plasmonic nanoaperture for ultrahigh optical force generation with minimal heat generation

Theodore Anyika[1,3], Ikjun Hong[1,3], Justus C. Ndukaife[1,2,3*]

[1]Department of Electrical and Computer Engineering, Vanderbilt University, Nashville, TN, USA

[2]Department of Mechanical Engineering, Vanderbilt University, Nashville, TN, USA

[3]Vanderbilt Institute of Nanoscale Science and Engineering

[4]Vanderbilt Center for Extracellular Vesicles Research

*justus.ndukaife@vanderbilt.edu

**Abstract**

Double Nanohole Plasmonic Tweezers (DNH) have revolutionized particle trapping capabilities, enabling trapping of nanoscale particles well beyond the diffraction limit. This advancement allows for the low-power trapping of extremely small particles, such as 20 $nm$ nanoparticles and individual proteins. However, to mitigate the potentially amplified effects of plasmonic heating at resonance illumination, DNH plasmonic tweezers are typically operated under off-resonance conditions. Consequently, this results in a decrease in optical forces and electric field enhancement within the plasmonic hotspot, which is undesirable for applications that require enhanced light-matter interaction like Surface Enhanced Raman Spectroscopy (SERS). In this study, we present a novel design for DNH plasmonic tweezers that addresses these limitations and provides significantly higher field enhancements. By introducing a reflector layer, on-resonance illumination can be achieved while significantly reducing plasmonic heating. This reflector layer facilitates efficient dissipation of heat both in-plane and axially. Furthermore, the integration of a reflector layer enables a redistribution of the hotspots via maxwell's boundary conditions for metals, creating more accessible hotspots optimal for applications that require enhanced light-matter interaction. We also demonstrate low-power trapping of small extracellular vesicles using our novel design, thereby opening possibilities for applications such as SERS and single photon emission that require intense light-matter interaction.

**Introduction**

The ability to precisely trap and manipulate micron/nanoscopic particles using optical tweezers has significantly impacted the field of particle manipulation and has found applications towards trapping and manipulating a wide range of materials, ranging from inorganic dielectric particles[1,2], to metal nanoparticles[3–5] and carbon nanotubes[6]. In biological research, optical tweezers have enabled several breakthroughs including, crucial nanoscale force measurements to determine the elastic properties of DNA strands[7], determining the refractive index of viruses using well calibrated optical traps[8], and in-situ single particle raman spectroscopy via raman tweezers micro-spectroscopy[9,10]. Despite the wide range of applications enabled by optical tweezers, the trapping and manipulation of nanoscopic particles requires very high laser powers due to the inherent diffraction limitation of these systems, which limits its applications in biological sciences to larger bio species like cells. The need for higher laser powers for nanoscale trapping is a consequence of the size-dependent nature of the optical gradient force, which necessitates stronger intensity gradients to trap smaller particles. By leveraging the highly intense electromagnetic hotspots created through the localized surface plasmon resonance (LSPR) of metallic nanostructures, plasmonic nanotweezers enable low-power trapping well beyond the diffraction limit, with ultra-narrow potential widths that ensure single particle trapping[11–17]. Although nanoscopic trapping beyond the diffraction limit has been demonstrated using plasmonic nanoantennas, positive thermophoretic effects arise due to the inherent non-radiative losses in metals which decrease the stability of the trap[11,17–19]. To circumvent unwanted heating effects in plasmonic antennas, the Crozier group demonstrated the trapping of a 110 $nm$ polystyrene particle using a resonance-tuned plasmonic nanopillar coupled with an integrated heat sink that comprised a stack of high thermal conductivity films serving as the substrate[20]. In contrast to plasmonic antennas, the use of nanoapertures in metallic films for nanoscale trapping effectively reduces plasmonic heating especially when illuminated off-resonance, as the continuous metallic film facilitates efficient heat dissipation[21]. Furthermore, the phenomenon of self-induced back action in nanopores, wherein the presence of a particle in a nanoaperture alters the surrounding electromagnetic field, thereby enhancing the trapping mechanism[11,22] makes them more suited for nanoscopic trapping. Double nanohole (DNH) apertures are particularly interesting owing to the highly enhanced electromagnetic hotspots in the 'gap' of the nanoaperture[14,21,23–25]. This presents exciting opportunities in biological sciences to trap and study single nanoscopic bio species. Pang et al.[26] trapped and observed the folding and unfolding of a single Bovine serum albumin (BSA) protein using a double nanohole aperture. Trapping of ~20 $nm$ [14,21,24] and ~12 $nm$ [23] dielectric beads have been demonstrated by various groups using DNH apertures in gold films. Despite the success of these nanoapertures, it is important to note that these systems are usually detuned from resonance to avoid substantial heat generation due to increased absorption of gold at longer wavelengths and the low thermal conductivity of the underlying glass substrate. Resonance-detuned operation results in less intense electromagnetic hotspots, which implies that these destined systems are not fully leveraging the benefits the plasmonic aperture to maximize the gradient force and enhance spectroscopy of taped specimens. Finite difference time domain (FDTD) simulations of the electromagnetic field distribution around DNH apertures show that the resonance detuned hotspots are mostly localized in the glass substrate[25], leaving the hotspots less accessible for applications requiring strong light-matter interaction like surface enhanced raman spectroscopy (SERS). This is due to the higher photon density of states in the glass substrate relative to that of the water supernatant. The trapping of single small extracellular vesicles (e.g. exosomes) holds great potential for investigating the heterogeneity within specific sub populations of these bio species, leveraging the intense electromagnetic hotspots for enhanced light-matter interaction. In this work, we present a novel DNH aperture design that simultaneously address the issue of less accessible hotspots and significant plasmonic heating for resonance tuned DNH apertures. To address these issues, we incorporate an integrated reflector layer into the DNH configuration. Additionally, we experimentally demonstrate low-power single small EV trapping using this novel DNH design.

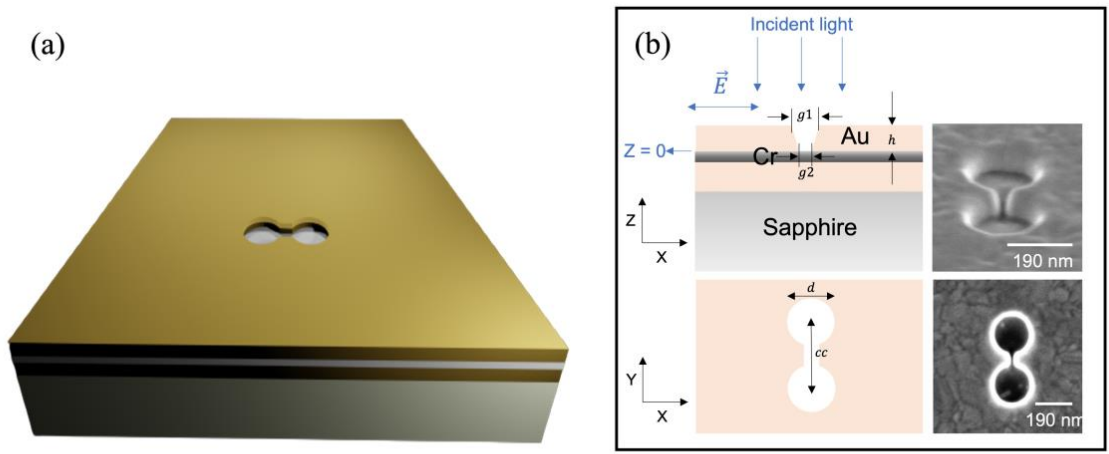

**Figure 1.** Schematic design for the double nanohole (DNH) on a reflector system. (a) Schematic design showing the integration of a Cr-Au hybrid reflector layer underneath the gold film. The Cr layer was selected for ease of fabrication and serves as a mask to prevent excessive drilling of the gold film layer during focused ion beam milling. (b) ZX and XY cross sections for the DNH illuminated under X polarized light. The widening of the DNH gap due to the focused ion beam milling fabrication process is shown as $g1$ while $g2$ is the bottom gap of the DNH. $d$ and $cc$ represent the diameter and the center-to-center distance of the nanoholes. The right panel shows scanning electron micrographs (SEM) of the DNH.

**Results and discussion**

The integration of a reflector layer directly beneath the Au film results in a redistribution of the electromagnetic hotspots around the top of the DNH gap. This redistribution of the electromagnetic field offers more accessibility to the hotspots for applications like SERS which involves enhanced light-matter interaction. Efficient heat dissipation away from the hotspots is achieved in-plane via the continuous Au film and axially via the high thermal conductivity substrate, comprised of the reflector layer deposited on a sapphire substrate with a higher thermal conductivity compared to the conventionally used silica substrate. The schematic design of this system is shown in figure 1(a) and (b), with left panel of figure 1(b) showing the SEM micrographs of DNH on a reflector system. The redistribution of the electromagnetic field due to the integration of the reflector layer results from maxwell's boundary condition for metals, which requires the electric fields within metals to tend towards zero. To investigate the properties of this new design, we first theoretically consider the conventional DNH configuration, an aperture in a $100\ nm$ thick Au film on a glass (silica) substrate. For this system, $d$ (diameter of each circle) is taken to be $100\ nm$, while $cc$ (which is the center-to-center spacing of the circles making the DNH) is taken as $140\ nm$. We have considered the effect of the top gap widening caused by the focused ion beam milling during sample fabrication. The top of the DNH gap $g1$ is taken to be $45\ nm$ while the bottom gap $g2$ is taken as $30\ nm$ for all simulations presented in this work. The incident electric field is polarized along the X axis as depicted in figure 1(b). The electromagnetic field enhancement distribution in the ZX plane at Y = 0 is shown for both resonance tuned and detuned scenarios in figures 2(a) and (b). The resonance detuned field enhancement is shown for a wavelength of $973\ nm$ which is the experimental trapping wavelength used throughout this work, while the on-resonance wavelength shown in figure 2(b) is $1265\ nm$.

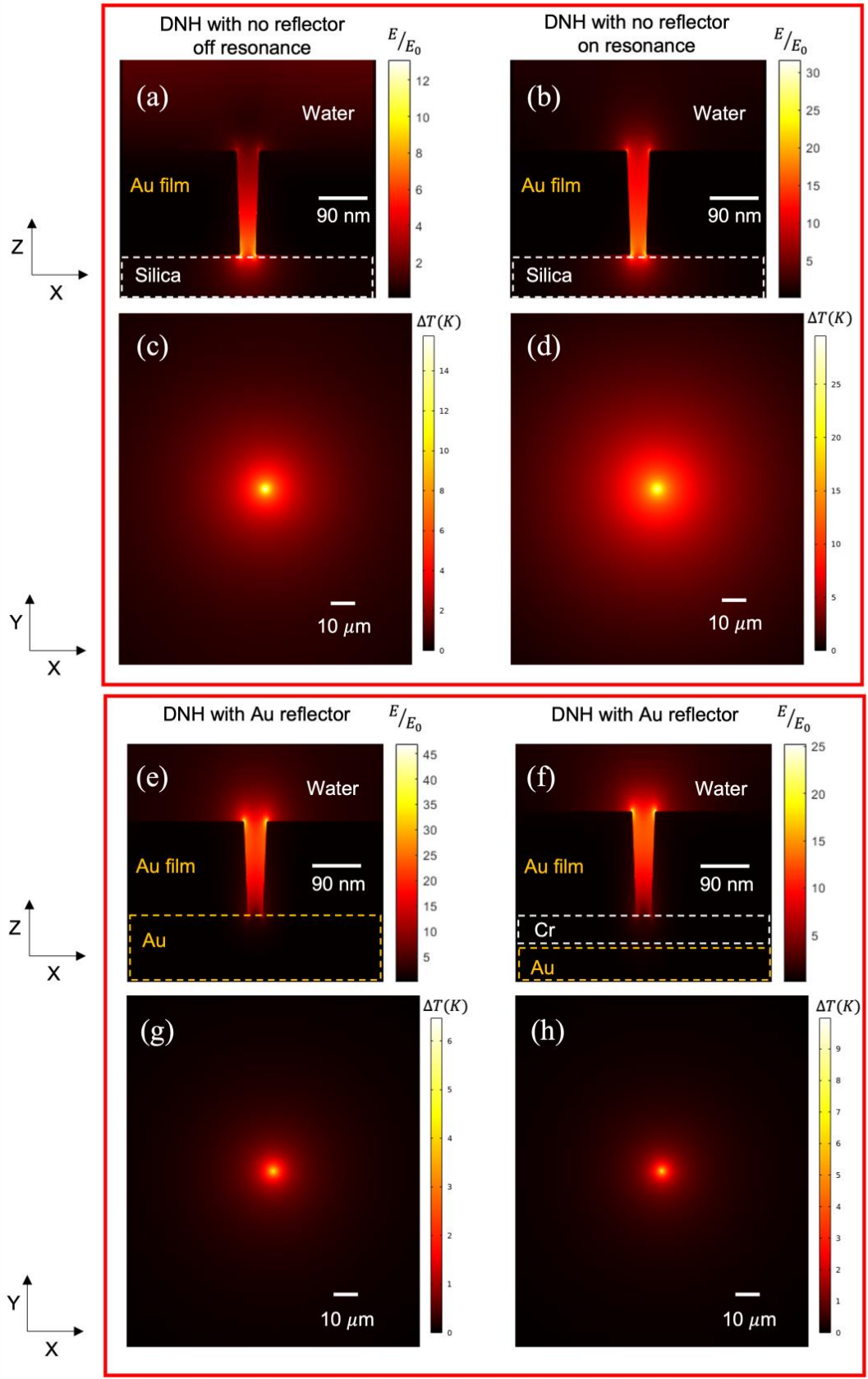

**Figure 2.** Comparative analysis of FDTD electromagnetic and FEM thermal simulations for X polarized incident light. (a) Electric field enhancement for the resonance detuned DNH on glass system in the ZX plane at X = 0, Y = 0, illuminated at 973 $nm$, while (b) shows the resonance tuned electric field enhancement distribution at the resonant wavelength of 1265 $nm$ in the ZX plane at X = 0, Y = 0. The nanohole diameter $d$ is taken to be 100 $nm$ while the center-to-center distance $cc$ between the nanoholes is taken as 140 $nm$ for figures (a) and (b). (c) and (d) show the corresponding FEM thermal simulations for figures (a) and (b) respectively in the XY plane at Z = 50 $nm$. (e) Electric field enhancement for a DNH with an integrated 150 $nm$ Au reflector layer in the ZX plane, at X = 0, Y = 0 (center of the DNH), tuned to be resonant at an incident wavelength of 973 $nm$. The top and bottom gaps $g1$ and $g2$ are taken to be 45 $nm$ and 30 $nm$ respectively. The nanohole diameter $d$ is taken to be 170 $nm$, the center-to-center distance $cc$ between the nanoholes is taken as 200 $nm$ while the Au film thickness is 100 $nm$. (f) The corresponding field enhancement distribution for a DNH in a 100 $nm$ Au film with a hybrid Cr-Au reflector layer tuned to the same resonance wavelength as (e), with the same values for $g1$ and $g2$. The nanohole diameter $d$ is taken to be 190 $nm$ while the center-to-center distance $cc$ between the nanoholes is taken as 260 $nm$. The hybrid reflector comprises a 50 $nm$ Cr layer and a 100 $nm$ Au layer. (g) and (h) show the corresponding FEM thermal simulations for figures (e) and (f) respectively in the XY plane at Z = 50 $nm$. The Yellow and white dotted lines show the distinction between various substrate materials.

For both resonance tuned and detuned scenarios, the structure parameters remain constant while only the illumination wavelengths change. The confinement of the hotspots in the glass substrate can be seen in figures 2(a) and (b), with a maximum field enhancement of 13 for the resonance detuned configuration as opposed to a maximum field enhancement of 31 for the resonance tuned configuration. This shows a reduction in the intensity enhancement by over a factor of 5 resulting from resonance detuning of the aperture. Furthermore, finite element method (FEM) simulations were run for both scenarios to investigate the plasmonic heating effect. The resonance tuned scenario in figure 2(d) shows a temperature rise of up to 30 $K$ while the resonance detuned case shows a 15 $K$ rise in temperature (figure 2(c)). Next, we theoretically investigate the resonance tuned DNH on a gold reflector system. For this system, we consider a DNH aperture in a 100 $nm$ thick Au film, on a 150 $nm$ Au reflector on a sapphire substrate. The combination of Au with a thermal conductivity of 317 $W/mK$ and sapphire with a thermal conductivity of 35 $W/mK$ results in efficient heat dissipation in the in-plane and axial directions, giving rise to minimal plasmonic heating effect for resonance tuned illumination. For this configuration, the nanohole diameter $d$ is taken to be 170 $nm$ while the center-to-center distance $cc$ is 200 $nm$. The gap parameters, $g1$ and $g2$ are taken as 45 $nm$ and 30 $nm$ respectively, while the incident light is polarized along the X axis. The resonance tuned electric field distribution in the ZX plane at Y = 0 is shown in figure 2(d). The redistribution of the electromagnetic field is apparent from the electric field enhancement distribution, showing more accessible hotspots. Additionally, we show that the maximum achievable electric field enhancement on resonance can be enhanced with the reflector layer by a factor of approximately 1.5 compared to the resonance tuned DNH on glass configuration. This results from the constructive interference between the reflected light and the gap plasmon modes in the nanogap. To investigate the temperature rise for this system, FEM thermal simulations were run as shown in figure 2(g). This shows an approximately 6.3 $K$ rise in temperature which is less than that of the resonance tuned DNH on glass by a factor of 5, despite having the most intense electric field hotspots. This is attributed to the efficient thermal engineering of the substrate. The DNH design with the integrated reflector layer not only provides more accessible hotspots, but also significantly minimized the temperature rise. The coupling of the reflected light to the gap plasmon mode for the DNH on a reflector configuration allows for easy tunability of the resonance for a given gap size by varying the thickness of the Au film, as opposed for the conventional DNH on glass configuration, which is more stringent with height variations, as shown in figures 3(b) and (d). This expands the parameter space for DNH plasmonic tweezers, making it easier to realize more efficient coupling to a 785 $nm$ raman excitation laser for SERS experiments. The disparity between the resonance and operation wavelengths commonly used in literature[21,24] for the DNH on glass system is clearly shown in Figure 3(c), underscoring the limitations associated with this off-resonance system. Additionally, Figure 3(d) shows limited tunability for this system beyond 1300 $nm$ , for a given gap size. To experimentally demonstrate nanoparticle trapping using the optimized DNH configuration, an important design consideration for the DNH reflector would be to include a hard mask layer to enable more accurate control of the nanohole depth during the focused ion beam milling process which is employed in this work. For this reason, we simulated the electromagnetic field distribution for the DNH on a reflector system, replacing the 150 $nm$ Au reflector layer with a resonance tuned hybrid Cr-Au reflector layer consisting of a 50 $nm$ Cr hard mask layer on a 100 $nm$ Au reflector. The diameter $d$ is given as 190 $nm$ while the center-to-center distance $cc$ is 260 $nm$. However, it is important to note that the maximum resonance tuned electric field enhancement for this hybrid reflector configuration is less than that of the Au reflector as shown in figure 2(f). This can be attributed to the damping of the LSPR resonance in the Au film by the Cr layer as seen from the normalized electric field enhancement plots in figure 3(a). Haykel *et al.*[27] showed that the incorporation of a chromium (Cr) adhesion layer within a plasmonic aperture leads to the attenuation and broadening of the LSPR resonance. Moreover, they found that an augmented thickness of the adhesion layer corresponds to greater damping effects. The FEM thermal simulation for the hybrid reflector system in the XY plane at Z = 50 $nm$ is presented in figure 2(h). This shows a maximum temperature rise of 10 $K$, which is higher than that obtained for the system with a uniform Au reflector. The higher temperature rise is attributed to the lower thermal conductivity of the Cr hard mask layer of 93 $W/mK$. This is shown in supplementary figure 1. The choice of a Cr hard mask layer used for experimental demonstration in this work is primarily due to fabrication considerations. To fabricate this device, a 10 $nm$ Cr adhesion layer was first deposited on a sapphire substrate via resistive deposition, followed by electron beam (e-beam) evaporation of a 100 $nm$ Au film. Next, the 50 $nm$ Cr hard mask was deposited resistively followed by a subsequent e-beam evaporation of a 100 $nm$ Au film. All processes were run in a multimode deposition chamber without breaking the vacuum. To define DNH on the Au film, we used a gallium source focused ion beam with a 1.1 $pA$ beam current that ablates the Au film much more efficiently compared to the Cr hard mask layer. Next, we treated the DNH sample using Poly (sodium 4-styrenesulfonate) (PSS) for 10 minutes, followed by a 5-minute rinse with potassium chloride (KCl). This process aimed to prevent the adsorption of negatively charged small EVs to the Au surface. PSS has been shown to passivate Au and quantum dot surfaces with negative charge[28,29]. Afterwards, the sample was blow-dried. Next, the DNH sample was made into a microfluidic chip, utilizing 120 $\mu m$ thick dielectric spacers and an ITO-coated cover slip to close the channel. Note that both the DNH sample and the ITO-coated cover slip underwent the same surface treatment. A 973 nm diode laser was utilized for trapping experiments. To ensure proper alignment, the polarization of the laser was oriented perpendicular to the center-to-center axis of the DNH, using alignment markers milled in the Au film. The laser was focused to achieve a spot size of 1.33 $\mu m$, utilizing a 40× 0.75 NA objective lens mounted on an inverted microscope.

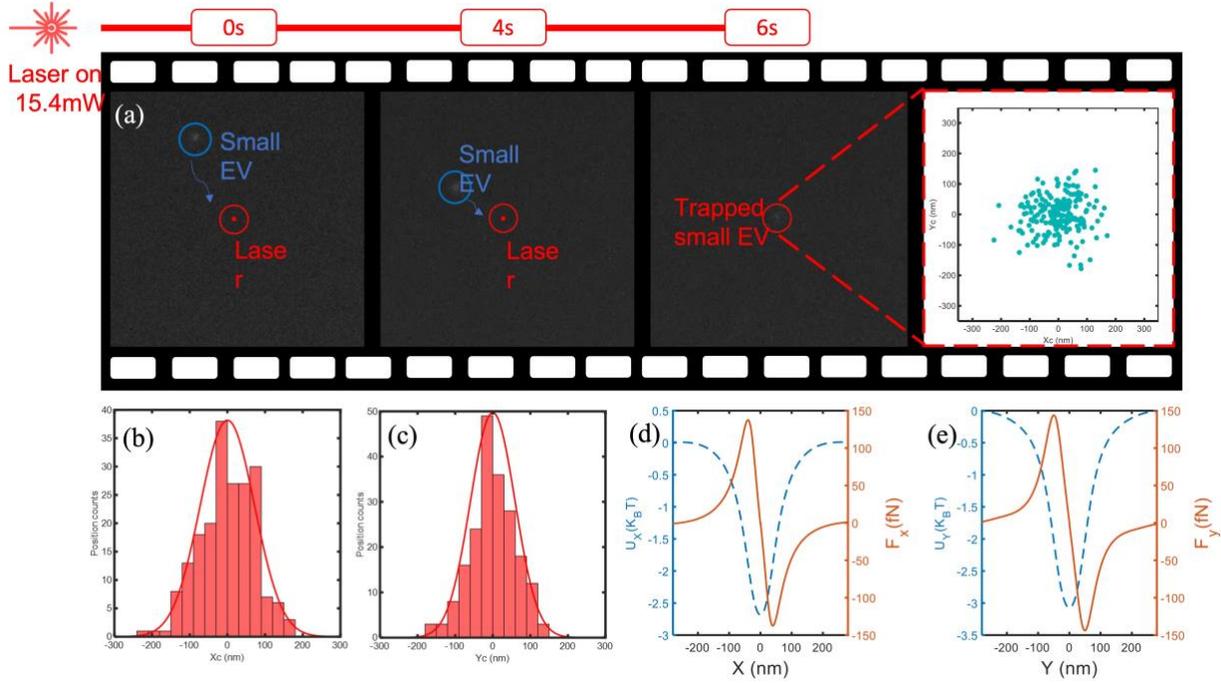

**Figure 4.** Experimental results and optical force calculations. (a) Experimental trapping video showing the diffusion of a small EV in the vicinity of the DNH and subsequent trapping (frames 1 - 3), with frame 4 shown the trapped particle displacement scatter plot, for a 15.4 $mW$ 973 $nm$ trapping laser. (b) and (c) show the particle displacement histograms along the X and Y axes respectively for the data in frame 4 of (a), with $X_c$ and $Y_c$ representing the center of the fluorescently labelled small EV. (d) and (e) show the X and Y components of the optical force (orange curves) exerted on a 100 $nm$ small EV, obtained using the Maxwell stress tensor (MST) method. The optical force calculations were performed by varying the particle center along the Y = 0 and X = 0 lines respectively, while maintaining the particle center position at Z = 55 $nm$. The corresponding optical potentials are shown as the blue dotted curves.

Commercially purchased neon green conjugated small EVs were suspended in deionized water and injected into the microfluidic chamber. Using a 15.4 $mW$ laser, we demonstrated trapping of small EVs as shown in figure 4(a). Frames 1 and 2 show a single small EV diffusing in the vicinity of the hotspot while frame 3 shows the particle trapped in place at the plasmonic hotspot. Using a CCD camera operating at a 200 $ms$ exposure time, we tracked the particle position of the fluorescently labelled small EV in real time using a custom python script. Supplementary movie 1 shows a small EV trapped under a 15.4 $mW$ laser, while the corresponding tracked particle canter displacement data over a duration of 30 seconds is given in frame 4 of figure 4(a). Histograms for the particle center displacement along the X and Y axes are shown in figures 4(b) and (c). To calculate the trap stiffness, we first obtain the experimental measured variance $Var_m(i)$ from the particle position displacement presented in figure 4(a). Using the equipartition theorem, the variance along the $ith$ axis is given by $Var(i) = \frac{K_B T}{k_i}$, where $K_B$ is the Boltzmann constant, $T$ is the temperature, $Var(i)$ is the variance and $k_i$ is the stiffness along the $ith$ axis. However, the experimentally obtained variance $Var_m(i)$ is bound to be less than the theoretical value due motion blur effects. To correct for this, we applied a motion blur correction function[30] given by $Var_m(i) = Var(i)\, S(\alpha)$, where $\alpha = \frac{tDk_i}{K_B T}$, $t$ is the exposure time and $D$ is the diffusion coefficient. The full form of $S(\alpha)$ as given in ref 28 was utilized to numerically solve for the experimental trap stiffness $k_i$. Using this approach, the X and Y trapping stiffnesses were determined to be $k_x = 0.0798\ fN/nm, k_y = 0.0974 fN/nm$. Previous works utilizing DNH apertures for trapping of 20 $nm$ polystyrene beads have shown trapping stiffnesses up to 0.2 $fN/nm$ [21,24]. However, it is important to note that small polystyrene beads have a refractive index of 1.6 which enhances the optical force compared to small EVs with a refractive index of ~1.39[31] Furthermore, these particles show good spatial overlap with the electromagnetic field distribution in the DNH gap. The relatively low trapping stiffness obtained from the small EVs can be attributed to the low overlap between the particle size and the DNH gap. Small EVs typically have a size distribution ranging from 30 – 200 $nm$ in diameter with a peak at around 100 $nm$. We show trapping and release videos in supplementary movies 2 and 3. To gain deeper insights into the process of particle trapping, we conducted calculations of the optical gradient force and potential. Specifically, we focused on a small EV (100 nm in size) and employed the Maxwell stress tensor method for our calculations. In these calculations, we assumed a refractive index of 1.39. Our analysis revealed that the optical potential obtained for the small EV was substantial. Figures 4(d) and (e) showcase the results, demonstrating that the optical potential energy reached values as high as $3K_B T$ in both the X and Y directions. These findings are significant as they indicate that the generated optical potential possesses sufficient energy to surpass the Brownian motion exhibited by the small EV. In other words, the trapping mechanism can effectively counteract the random thermal motion of the particle, enabling stable confinement within the trap.

**Conclusion**

The integration of a reflector layer in the double nanohole (DNH) aperture design offers significant advantages for nanoscale trapping and manipulation. The incorporation of the reflector layer redistributes the electromagnetic hotspots, making them more accessible for applications such as surface-enhanced Raman spectroscopy (SERS) that require strong light-matter interaction. Additionally, the reflector layer enables efficient heat dissipation, further minimizing plasmonic heating effects. The optimized DNH configuration demonstrates low-power trapping of small extracellular vesicles (exosomes) with improved field enhancement and reduced temperature rise. The experimental demonstration of particle trapping shown here using the DNH reflector system validate its effectiveness, expanding the possibilities for more efficient light-matter interaction applications in areas such as biological sciences and surface-enhanced Raman spectroscopy. The ability to trap single small extracellular vesicles using the highly intense electromagnetic hotspots of the DNH aperture holds great promise for single small EV analysis using Raman spectroscopy. In recent studies, the potential of small extracellular vesicles (EVs) has been highlighted as biomarkers for the detection of cancer. However, the practical application of small EVs in clinical settings for early cancer detection is limited by the considerable heterogeneity observed among these biological species[32–34]. The optimized DNH on a reflector design proposed in this work allows for more accessible hotspots with minimal heating effects, which can facilitate the investigation of the heterogeneity exhibited by these biological species at the individual particle level via SERS, thereby paving the way for a more comprehensive understanding of their characteristics.

**Acknowledgements**

We acknowledge financial support from the National Science Foundation (NSF) CAREER Award (NSF ECCS 2143836).

## Supplementary Information

## Thermal calculations

Using COMSOL Multiphysics, we solved equation (1) for the electric field $\mathbf{E}$ in the vicinity of the DNH,

$$\nabla \times \nabla \times \mathbf{E} - k_0^2 \varepsilon(r)\mathbf{E} = 0 \tag{1}$$

where $k_0$ is the wave number given by $2\pi/\lambda$ and $\varepsilon(r)$ is the complex permittivity of the medium. Next, the power dissipation density in a $1064 \times 1064\ nm$ bounding region in the Au film around the DNH was determined from the equation, $q(\mathbf{r}) = 1/Re(\mathbf{J} \cdot \mathbf{E})$, where $\mathbf{J}$ is the current density induced in the vicinity of the plasmonic aperture. The heat dissipation density in the rest of the Au film was computed from equation (2),

$$q(r) = P_0 A_{au} \frac{\alpha_{au}}{\pi \sigma^2} e^{-(\frac{r^2}{2\sigma^2})} e^{-\alpha_{au} z} \tag{2}$$

where, where $P_0$ is the incident power, $A_{au}$ is the absorption of the Au film, $\alpha_{au}$ is the absorption coefficient of the Au film at the illumination wavelength. The first exponential term in equation (2) considers the influence of the Gaussian beam, with $\sigma$ representing the waist radius of the focused laser. On the other hand, the second exponential term considers Beer's law. Lastly, the temperature rise was then computed by solving the heat transfer equation (3),

$$-k\nabla^2 T + \rho C_P u \cdot \nabla T = q(r) \tag{3}$$

where, $k$ is the thermal conductivity of the materials, $T$ is the temperature in Kelvin, $\nabla T$ is the temperature gradient, $C_P$ is the specific heat capacity at constant pressure, $\rho$ is the density of the fluid, $u$ is the fluid velocity (taken as zero in the Au film and reflector layers), and $q(r)$ is the heat source density.

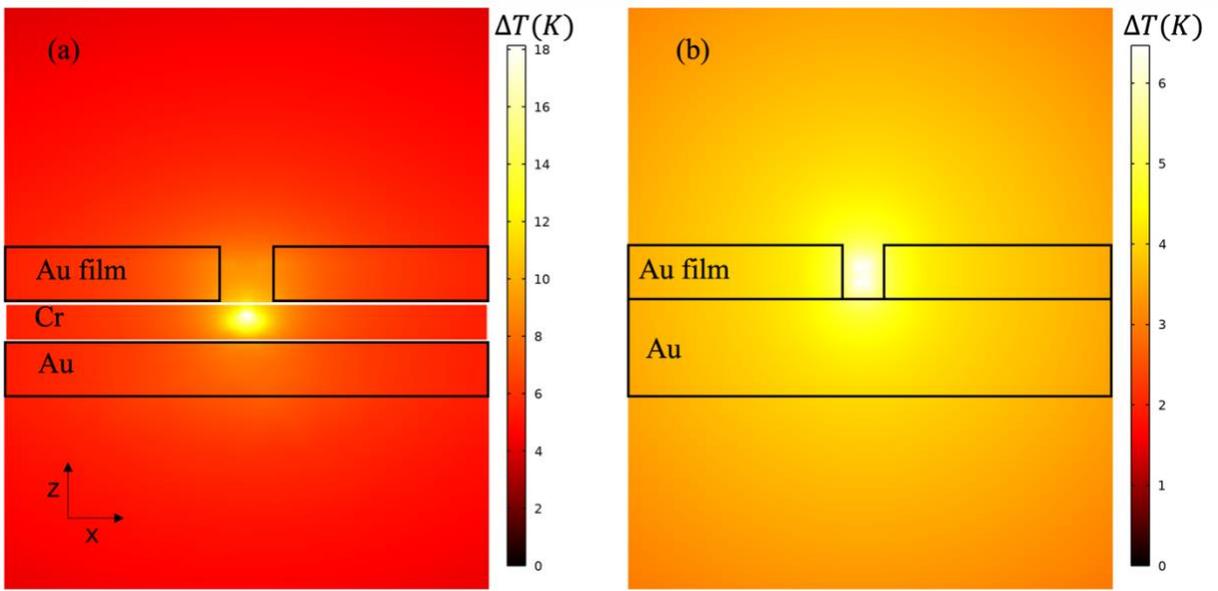

**Supplementary Figure 1.** (a) Temperature rise distribution in the ZX plane at Y = 0 for the DNH on a Cr-Au reflector showing higher temperature rise in the Cr layer due to the lower thermal conductivity of Cr relative to Au. (b) DNH on a uniform Au reflector showing less temperature rise.

**Movie 1:**  small EV trapped under a 15.4 $mW$ laser.
**Movie 2:**  Transport and trapping of a small EV.
**Movie 3:**  Release of a small EV after the laser was turned off.